\documentclass[11pt]{article}
\usepackage{epsfig}

\newcommand{\MeV}{\mbox{\rm\,MeV}}
\newcommand{\GeV}{\mbox{\rm\,GeV}}
\newcommand{\pfrac}[2]{\left(\frac{#1}{#2}\right)}
\newcommand{\artanh}{\mathop{\rm artanh}\nolimits}

\begin{document}
\thispagestyle{empty}
\begin{flushright}
MZ-TH/01-31\\
hep-ph/0111206\\
November 2001
\end{flushright}
\vspace{0.5cm}
\begin{center}
{\Large\bf An interpolation of the vacuum polarization function\\[.3truecm]
  for the evaluation of hadronic contributions\\[.4truecm]
  to the muon anomalous magnetic moment}\\[1truecm]
{\large \bf S.~Groote, J.G.~K\"orner and A.A.~Pivovarov}\\[.7truecm]
Institut f\"ur Physik, Johannes-Gutenberg-Universit\"at,\\
Staudinger Weg 7, D-55099 Mainz, Germany\\[.3truecm] and \\[.3truecm]
Institute for Nuclear Research of the\\
Russian Academy of Sciences, Moscow 117312, Russia
\end{center}

\vspace{1truecm}
\begin{abstract}\noindent
We propose a simple parameterization of the two-point correlator of hadronic
electromagnetic currents for the evaluation of the hadronic contributions to
the muon anomalous magnetic moment. The parameterization is explicitly done in
the Euclidean domain. The model function contains a phenomenological parameter
which provides an infrared cutoff to guarantee the smooth behavior of the
correlator at the origin in accordance with experimental data in $e^+e^-$
annihilation. After fixing a numerical value for this parameter from the
leading order hadronic contribution to the muon anomalous magnetic moment the
next-to-leading order results related to the vacuum polarization function are
accurately reproduced. The properties of the four-point correlator of hadronic
electromagnetic currents as for instance the so-called light-by-light
scattering amplitude relevant for the calculation of the muon anomalous
magnetic moment are briefly discussed.
\end{abstract}

\newpage
\section{Introduction}
The description of strong interactions based on QCD proves to be very
successful for processes at large energies where the coupling constant is
small due to the property of asymptotic freedom~\cite{Gross:1973id}. This
makes perturbation theory (PT) computations reliable. At low energies the
problem of strong coupling prevents using QCD as an unambiguous theoretical
tool for computations of physical observables and various phenomenological
models are introduced. These models are inspired by QCD but it is difficult to
establish a quantitative relation between the underlying theory and a model
used in practice. For instance, the chiral perturbation theory (ChPT) for
Goldstone modes is a very general model in the sense that this effective
theory provides the expansion of the full QCD amplitudes at low energies
according to symmetry principles without using special assumptions. In this
respect the amplitudes of ChPT represent the low energy limits of the exact
QCD amplitudes based on current algebra and collect the current
algebra results in a compact way~\cite{Weinberg:1967fm}. Of course, for
the applicability of ChPT one should guarantee that the expansion in the
energy is well convergent (at least explicitly). ChPT is very practical
in describing interactions of pions (as lightest hadrons) with nucleons or
resonances at low energy in the small momentum (and mass)
expansion~\cite{phenlagr,Gasser:1984yg}. Thus the description of strong
interactions at low energies relies on phenomenological models with explicit
introduction of elementary hadron fields or on the closely related approaches
based on general principles of analyticity, unitarity and symmetry~\cite{chew}.
A general idea of linking this approach for the description of hadrons at low
energies with QCD is the concept of duality which means that the description
of inclusive observables which are sensitive to the contribution of many
particles is simpler than that of exclusive processes and can be represented
by almost free fermions or weakly coupled quarks~\cite{Poggio:1976af}. This
concept works well for infrared (IR) soft observables in $\tau$-decays and
other sum rules where the limit of massless quarks is
nonsingular~\cite{Shif,Kra1983,reinders,Bra1992qm,Piv1992rh0,Col2000dp}. For
the IR sensitive observables the realization of the duality concept for the
light modes is not quite straightforward since the IR cutoff explicitly enters
the calculation. In such cases the IR cutoff is usually taken from experiment
such as the mass of a real hadron.

Strong interactions at low energies play an important role in precision tests
of the Standard Model as a whole. They are involved in the evaluation of the
CP-violating structure in the electroweak sector through hadronic matrix
elements, the computation of which constitutes a main obstacle for the
progress of a quantitative analysis of the nonleptonic kaon decays relevant
for determination of the quark mixing matrix~\cite{buras,penpivO8} and mixing
of neutral pseudoscalar states~\cite{kk}. In other tests of the Standard Model 
strong interactions enter as small corrections to very accurately measured
observables. Examples of high precision observables of the Standard Model with
important contributions from strong interactions (as corrections to the
leading leptonic effects) are the running electromagnetic (EM) coupling
constant at the scale of the $Z$-boson mass and the muon anomalous magnetic
moment (MAMM). The numerical values of these quantities put some constraints
on the Standard Model parameters and can also serve as triggers of new physics
beyond the Standard Model.

A numerical value of the muon anomalous magnetic moment (MAMM) is measured
experimentally with high precision~\cite{PDG,expamu}. The value presented in a
recent review~\cite{czamar} reads 
\begin{equation}
\label{exp}
a_\mu^{\rm exp}=116~592~023(151)\times 10^{-11}
\end{equation}
with an uncertainty of $151\times 10^{-11}$. In future experiments a goal is
set to reach the accuracy of $40\times 10^{-11}$. In theoretical computations
the leading contribution to the MAMM is given by
\begin{equation}
\label{Schw}
a_\mu^{\rm Schw}=\frac{\alpha}{2\pi}
\end{equation}
(first calculated by Schwinger) where $\alpha$ is the fine structure constant 
$\alpha^{-1}=137.036\dots$~\cite{PDG}. The MAMM is sensitive to the IR region
of integration in perturbation theory diagrams. For the leptonic (QED) part of
the contribution this is reflected in a strong dependence on the electron
mass. Theoretically the purely leptonic part is computed in perturbative QED
with finite lepton masses which leads to a function
$A(m_e/m_\mu, m_\mu/m_\tau)$ which is known analytically to three loops. As
$m_e\ll m_\mu \ll m_\tau$ the ratios $m_e/m_\mu$ and $m_\mu/m_\tau$ are small
and the function $A(m_e/m_\mu, m_\mu/m_\tau)$ can be expanded in these ratios
to simplify calculations. The contribution of the muon leads to diagrams with
a single scale $m_\mu$ that makes it simpler to compute. The complete
analytical calculation is technically very complicated already at the level of
three loops~\cite{lblanal}. The nontrivial diagrams in higher orders were
computed numerically~\cite{tenthQED,qedcont}. The present value of the QED
contribution to the muon anomalous magnetic moment
reads~\cite{pseudo,Kinoshita:1996vz} (as a review see~\cite{czamar})
\begin{equation}
\label{QED}
a_\mu^{\rm QED}=116~584~705.7(2.9)\times 10^{-11}.
\end{equation}
Computation of the electroweak (EW) corrections to the MAMM has also been
performed in perturbation theory~\cite{Brodsky:1968sr}. The EW contribution is
now known with two-loop accuracy (as a review see~\cite{czamar}),
\begin{equation}
\label{EW}
a_\mu^{\rm EW}=152(4)\times 10^{-11}. 
\end{equation}
The hadronic contribution to the MAMM is sensitive to the infrared region and
cannot be computed in perturbative QCD with light quarks. The current masses
of light quarks are too small to provide a necessary infrared cutoff and
explicit models of hadronization are required for a quantitative
analysis~\cite{Gasser:1982ap,maltman,Kataev:1983xu}. The hadronic contribution
is the main uncertainty in the theoretical computation of the MAMM in the
Standard Model. Assuming the validity of the Standard Model for the
description of elementary particle interactions
\begin{equation}
\label{assumption}
a_\mu^{\rm exp}=a_\mu^{\rm SM},
\end{equation}
the numerical value for the hadronic contribution to the MAMM in the Standard
Model is given by
\begin{equation}
\label{th}
a_\mu^{\rm had}|_{\rm as}=a_\mu^{\rm exp}-a_\mu^{\rm QED}-a_\mu^{\rm EW}
=(7165.3\pm 151|_{\rm exp} \pm 2.9|_{\rm QED}\pm 4|_{\rm EW})\times 10^{-11}.
\end{equation}
The experimental error dominates the uncertainty.

Since the hadronic contribution is sensitive to the details of the strong
coupling regime of QCD at low energies and cannot be unambiguously computed in
a perturbation theory framework the theoretical prediction for the MAMM in the
Standard Model depends crucially on how this contribution is
estimated~\cite{kinohad}. In the absence of a reliable theoretical tool for
the computation in this region one turns to experimental data on low-energy
hadron interactions for extracting a numerical
value~\cite{Jegerlehner:2001wq}. In general terms the hadronic contribution to
the MAMM is determined by the correlation functions of electromagnetic (EM)
currents. Since a source for the EM current is readily available for a wide
range of energies, one tries to extract these functions or some of their
characteristics relevant for a particular application from experiment. Without
explicit use of QCD the correction $a_\mu^{\rm had}$ in the Standard Model is
generated through the EM interaction $e j_\mu^{\rm had}A^\mu$ with
$j_\mu^{\rm had}$ being the hadronic part of the EM current. At the leading
order ($\alpha^2$ in formal power-counting) only the two-point correlation
function of the EM currents emerges in the analysis of hadronic contributions
to the MAMM. At the next-to-leading order ($\alpha^3$) a four-point
correlation function appears. These correlators are not calculable
perturbatively in the region essential for the determination of the hadronic
contributions to the MAMM. The leading contribution to the MAMM comes from the
two-point correlator referred to as the hadronic part of the photon vacuum
polarization contribution while the four-point function first emerges at the
$\alpha^3$ order, most explicitly as the light-by-light scattering amplitude.
To avoid using QCD in the strong coupling mode one can extract the necessary
contribution to the MAMM by studying these two correlation functions
experimentally without an explicit realization of the hadronic EM current
$j_\mu^{had}$ in terms of elementary fields. Another possibility which is
close in spirit is to use phenomenological models to saturate these
correlators with contributions of real hadrons at low
energies~\cite{pseudo,sakurai,mdif,Hayakawa:1996ki,Bijnens:1996xf}. There is
also a possibility to use a concept of duality between hadron and quark-gluon
descriptions modified for handling IR sensitive observables~\cite{eucl,anom1}.
In the following we discuss this last option.

\section{Hadronic contribution at leading order}
At the leading order in $\alpha$ the hadronic contribution is described by the
correlator 
\begin{equation}
\label{vacpolin}
i\int \langle Tj_\mu^{had}(x)j_\nu^{had}(0) \rangle e^{iqx}dx=
(q_\mu q_\nu -g_{\mu\nu}q^2)\Pi^{\rm had}(q^2)
\end{equation}
in terms of a single function $\Pi^{\rm had}(q^2)$ of one variable $q^2$. The
contribution of $\Pi^{\rm had}(q^2)$ to the muon anomalous magnetic moment
(e.g.\ \cite{leadorder}) is given by
\begin{equation}
\label{directcont}
a_\mu^{\rm had}({\rm LO})
=4\pi\pfrac{\alpha}{\pi}^2\int_{4m_\pi^2}^\infty
\frac{ds}{s}K(s){\rm Im}~\Pi^{\rm had}(s)
\end{equation}
with a one-loop kernel of the form
\begin{equation}
\label{oneloopkern}
K(s)=\int_0^1dx\frac{x^2(1-x)}{x^2+(1-x)s/m^2}.
\end{equation}
Here ${\rm Im}~\Pi^{\rm had}(s)={\rm Im}~\{\Pi^{\rm had}(q^2)|_{q^2=s+i0}\}$,
$m$ is a muon mass.
\begin{figure}
\centerline{\epsfig{figure=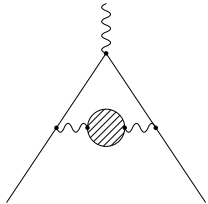, scale=0.2}}
\caption{\label{fig1}The leading order hadronic contribution to the MAMM,
  the shaded bubble indicates the hadronic two-point correlator}
\end{figure}

The leading order hadronic contribution to the MAMM as depicted in
Fig.~\ref{fig1} is represented by an integral over the hadron spectrum and no
specific information about the function ${\rm Im}~\Pi^{\rm had}(s)$ is
necessary point-wise. However, a QCD approach based on light quark duality in
the massless approximation is not directly applicable as the integral in
Eq.~(\ref{directcont}) is IR sensitive and depends strongly on the threshold
structure of the function $\Pi^{\rm had}(q^2)$. In most applications the
threshold structure is extracted from experiment. To the leading order in
$\alpha$ the function ${\rm Im}~\Pi^{\rm had}(s)$ can uniquely be identified
with data from $e^+e^-$ annihilation into hadrons. Introducing the
experimental $R^{\rm exp}(s)$ ratio
\begin{equation}
\label{expR}
R^{\rm exp}(s)
=\frac{\sigma(e^+e^-\to {\rm hadrons})}
{\sigma(e^+e^-\to \mu^+\mu^-)},
\qquad s=(p_{e^+}+p_{e^-})^2
\end{equation}
and identifying it with the theoretical quantity
$12\pi{\rm Im}~\Pi^{\rm had}(s)$ one finds
\begin{equation}
\label{exp0vac}
a_\mu^{\rm had}({\rm LO})
=\frac{1}{3}\pfrac{\alpha}{\pi}^2\int_{4m_\pi^2}^\infty
\frac{R^{\rm exp}(s)K(s)}{s}ds.
\end{equation}
The contribution to the MAMM based on the representation given in
Eq.~(\ref{exp0vac}) is well studied. Several recent determinations based on a
thorough treatment of various sets of data are given in the second column
of Table~\ref{tab1}.
\begin{table}[t]\begin{center}
\begin{tabular}{|r|l|l|l|l|}\hline
  reference&$a_\mu^{\rm had}({\rm LO})$&$a_\mu^{\rm had}({\rm LO;light})$
    &$m_{\rm eff}$ [\MeV]&$m_q$ [\MeV]\\\hline
  Ref.~\cite{kinohad}&$7011(94)\times 10^{-11}$
    &$6940(96)\times 10^{-11}$&$200.5(1.6)$&$178.7(1.4)$\\
  Ref.~\cite{jeger}&$7024(154)\times 10^{-11}$
    &$6953(155)\times 10^{-11}$&$200.3(2.5)$&$178.5(2.2)$\\
  Ref.~\cite{davhock1}&$6950(150)\times 10^{-11}$
    &$6879(151)\times 10^{-11}$&$201.5(2.5)$&$179.6(2.2)$\\
  Ref.~\cite{davhock2}&$6924(62)\times 10^{-11}$
    &$6853(65)\times 10^{-11}$&$201.9(1.1)$&$180.0(1.0)$\\
  Ref.~\cite{Jegerlehner:2001wq}&$6988(111)\times 10^{-11}$
    &$6917(112)\times 10^{-11}$&$200.9(1.8)$&$179.0(1.6)$\\
\hline
  optimistic&$6961(43)\times 10^{-11}$
    &$6892(44)\times 10^{-11}$&$201.3(0.7)$&$179.4(0.6)$\\
  conservative&$6979(114)\times 10^{-11}$
    &$6908(116)\times 10^{-11}$&$201.0(1.9)$&$179.2(1.7)$\\
\hline
\end{tabular}\end{center}
\caption{\label{tab1}Comparison of a selection of different determinations of
  the leading order contribution to $a_\mu^{\rm had}$ in the literature. For
  the first five lines the first column indicates the reference, the second
  column the cited value. In the third column the leading order contributions
  from heavy quarks are subtracted. The sixth and seventh line show the
  optimistic and conservative estimate for the mean value. The fourth and
  fifth column list the resulting values for $m_{\rm eff}$ and $m_q$ for
  model~1 and model~2, resp.}
\end{table}
All values given in Table~\ref{tab1} are consistent within errors. We can
take only a small selection of results because other results are again within
errors but with mean values slightly different. For the same reason, the old
historical result given in Ref.~\cite{kinohad} could be cited as well.
Table~\ref{tab1} gives two different mean values, namely an optimistic and
a conservative average. The optimistic average uses the error estimates of
the different results as weights, assuming that the results are independent.
Here we use the method presented in Sec.~4.3 of Ref.~\cite{jeger}. But because
all treatments in the literature are mainly based on the close data sets taken
for the analysis, this assumption might not be true. Therefore, we also give a
conservative average which reads
\begin{equation}
\label{piv}
a_\mu^{\rm had}({\rm LO})=6979(114)\times 10^{-11}.
\end{equation}
This average is in agreement with the experimental result in Eq.~(\ref{th})
within error bars. The statistical correlation of errors coming from the
experimental value in Eq.~(\ref{exp}) and the leading order hadronic data
in Eq.~(\ref{piv}) is supposed to be small as they are determined by different
sources. Other errors are negligible. For the target experimental error of the
MAMM at the level of $40\times 10^{-11}$~\cite{pseudo} the value in
Eq.~(\ref{piv}) is somewhat small and NLO hadronic corrections should be taken
into account. In order to obtain a naive order-of-magnitude estimate for
this contribution, we take the value of the leading order suppressed by
$\alpha\approx 1/137$ which gives roughly $50\times 10^{-11}$. Writing 
\[
a_\mu^{\rm had}|_{\rm th}= a_\mu^{\rm had}({\rm LO})
+ a_\mu^{\rm had}({\rm NLO})
\]
and comparing with Eq.~(\ref{th}) one has (in units of $10^{-11}$)
\begin{eqnarray}
a_\mu^{\rm had}({\rm NLO})
  &=&7165\pm 151|_{\rm exp}\pm 2.9|_{\rm QED}\pm 4|_{\rm EW}
  -6979(114)|_{\rm LO}\nonumber\\[7pt]
  &=&186\pm 151|_{\rm exp}\pm 2.9|_{\rm QED}\pm 4|_{\rm EW}\pm 114|_{\rm had}.
\end{eqnarray}
Assuming the statistical independence of the uncertainties one finds after
adding them in quadratures
\begin{equation}
\label{nlopres}
a_\mu^{\rm had}({\rm NLO}) = (186\pm 189)\times 10^{-11}
\end{equation}
which would not allow one to see higher order hadronic effects clearly.
Assuming that the mean value of $a_\mu^{\rm exp}$ in the planned experiment
will not change and the target accuracy $40\times 10^{-11}$ will be reached,
one finds a numerical value for the NLO hadronic contribution of
\begin{equation}
\label{futureNLOexp} 
a_\mu^{\rm had}({\rm NLO}) = (186\pm 121)\times 10^{-11}
\end{equation}
which makes the NLO hadronic effects noticeable at the level of two standard
deviations. If the mean value of $a_\mu^{\rm exp}$ will change in the range of
the present experimental uncertainty of $151\times 10^{-11}$, the NLO hadronic
effects can be more or less pronounced.

To create a framework for the analysis of hadronic contributions at NLO based
on duality arguments we rewrite the LO expression for the hadronic
contribution to the MAMM given in Eq.~(\ref{directcont}) in a different form.
From general principles a two-point correlator $\Pi(q^2)$ as a function of the
complex variable $q^2$ can have a cut along the positive semiaxis $s>0$ with a
positive discontinuity~\cite{Lehmann:1954xi}. This spectral condition plays
a crucial role in the analysis of the structure of the two-point correlators
and related observables~\cite{Krasnikov:1980yg,spect}. The dispersion
relation reads
\begin{equation}
\label{disprelgen}
\Pi(q^2)=\frac1\pi\int_0^\infty\frac{ds}{s-q^2}
  {\rm Im}~\Pi(s)-{\rm subtractions}.
\end{equation}
For massive pions the experimental spectrum in $e^+e^-$ annihilation starts
from $4m_\pi^2$ and the subtraction at the origin is possible as
${\rm Im}~\Pi^{\rm had}(s)=0$ for $s<4m_\pi^2$. The dispersion representation
with subtraction at the origin reads
\begin{equation}
\label{disprelhad}
\Pi^{\rm had}(q^2)
=\frac{q^2}{\pi}\int_{4m_\pi^2}^\infty
\frac{ds}{s}\frac{{\rm Im}~\Pi^{\rm had}(s)}{s-q^2}
\end{equation}
which implies the normalization condition $\Pi^{\rm had}(0)=0$. Using
Eqs.~(\ref{disprelgen},\ref{disprelhad}) and
Eqs.~(\ref{directcont},\ref{oneloopkern}) one can rewrite the LO contribution
to the MAMM as an integral over Euclidean values of $q^2$ for
$\Pi^{\rm had}(q^2)$,
\begin{equation}
\label{directcontEucl}
a_\mu^{\rm had}({\rm LO})=4\pi^2\pfrac{\alpha}{\pi}^2
\int_0^\infty \left\{-\Pi^{\rm had}(-t)\right\} W(t)dt
\end{equation}
with
\begin{equation}
W(t)=\frac{4m^4}{\sqrt{t^2+4m^2t}\left(t+2m^2+\sqrt{t^2+4m^2t}\right)^2}.
\end{equation}
Such a representation is well known and is often written as a parametric
integral~\cite{raf2,samuel}.

The representation in Eq.~(\ref{exp0vac}) is suitable for the evaluation of
the hadronic contributions to the MAMM by using experimental data, since it
can be rewritten in terms of the hadronic cross section for $e^+e^-$
annihilation. The representation in Eq.~(\ref{directcontEucl}) is more
suitable for a theoretical study as perturbation theory should preferably be 
applied in the Euclidean domain. Integration by parts in
Eq.~(\ref{directcontEucl}) gives
\begin{equation}
\label{ffuncparts}
\frac{1}{\pi}\int_{4m_\pi^2}^\infty
\frac{ds}{s}K(s){\rm Im}~\Pi^{\rm had}(s)=
\int_0^\infty \left(-\frac{d \Pi^{\rm had}(-t)}{d t}\right) F(t)dt,
\qquad
F(t)=\int_t^\infty W(\zeta)d\zeta
\end{equation}
with
\begin{equation}
\label{expliffun}
F(t)=\frac{1}{2}\pfrac{t+2m^2-\sqrt{t^2+4m^2t}}{t+2m^2+\sqrt{t^2+4m^2t}}
  =\frac{2m^4}{\left(t+2m^2+\sqrt{t^2+4m^2t}\right)^2}.
\end{equation}
\begin{figure}\begin{center}
\epsfig{figure=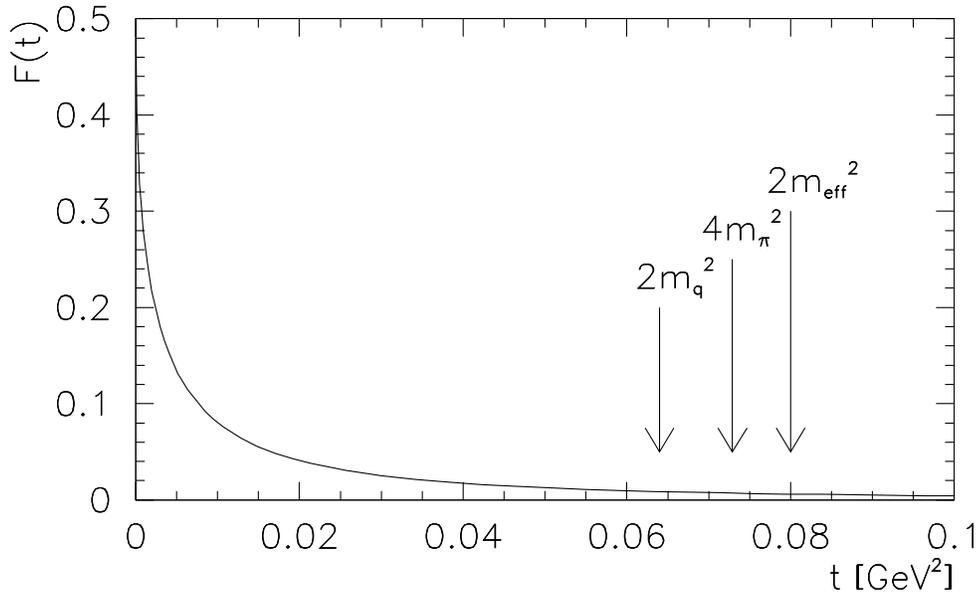, scale=0.8}
\caption{\label{fig2}The LO Euclidean weight function $F(t)$}
\end{center}\end{figure}
The behavior of the function $F(t)$ is shown in Fig.~\ref{fig2} for small and
large $t$. It reads
\begin{equation}
F(t)|_{t\to 0}=\frac{1}{2}-\frac{\sqrt{t}}{m}+O(t),
\qquad
F(t)|_{t\to \infty}=\frac{m^4}{2 t^2}+O(1/t^3).
\end{equation}
The surface terms of the integration by parts vanish because the integrand in
Eq.~(\ref{directcontEucl}) satisfies the conditions
$|\Pi^{\rm had}(-t)|< C \sqrt{t}$ at small $t$ and
$|\Pi^{\rm had}(-t)|< C^\prime t^{-2}$ at large $t$ with some 
given constants $C$, $C^\prime$.

A key physical quantity of the analysis is the derivative of the hadron vacuum
polarization function $d\Pi^{\rm had}(-t)/dt$ which is closely related to the
Adler function
\begin{equation}
D(t)= -t\frac{d \Pi^{\rm had}(-t)}{dt}.
\end{equation}
The quantity $d \Pi^{\rm had}(-t)/dt$ can be computed in perturbative QCD with
massless quarks for large $t$,
\begin{equation}
-\frac{d \Pi^{\rm had}(-t)}{dt}=\frac{e_q^2N_c}{12\pi^2t}
  \left(1+\frac{\alpha_s(t)}{\pi}\right)
\end{equation}
where $e_q$ is the charge of the quark in units of the elementary (electron)
electric charge and $N_c$ is the number of colors. Computation at small $t$ in
perturbation theory is not possible for light quarks with small masses as the
theory enters the regime of strong coupling. The behavior of the function
$d\Pi^{\rm had}(-t)/dt$ for small $t$ can be extracted from experiment where
the lower limit of the spectrum is determined by the finite pion masses. This
leads to a finite value for the function $d\Pi^{\rm had}(-t)/dt$ at $t=0$.
Using the patterns of small and large $t$ behavior of the function
$d\Pi^{\rm had}(-t)/dt$ for the light modes, we suggest an interpolation
function $f(t)$ valid for all $t$ in the form
\begin{equation}
\label{expint}
-\frac{d \Pi^{\rm had}(-t)}{dt}=\frac{e_q^2N_c}{12\pi^2}f(t),\qquad
  f(t)=\frac1{t+\Delta}.
\end{equation}
Writing
\begin{equation}
f(t)=-\frac{d p(t)}{dt}
\end{equation}
one has 
\begin{equation}
\label{modstep}
p(t)= \ln\pfrac{\Delta}{t+\Delta},
\qquad p(0)=0.
\end{equation}
The analytic properties of the function $p(t)$ are given by the cut along the
positive semiaxis starting at $s=\Delta$. The discontinuity across the cut is
equal to one,
\begin{equation}
\label{stepf}
r(s)=\frac1\pi{\rm Im}~p(-s-i0)=\theta(s-\Delta).
\end{equation}
Thus the contribution to the MAMM contains an integral
\begin{equation}
\label{integ}
I(\Delta) = \int_0^\infty f(t)F(t)dt
\end{equation}
which is the basic quantity for the theoretical analysis. The analytical
expression for $I(\Delta)$ is available but too cumbersome to be presented
here. This expression is used in numerical calculations. However, in order to
understand the integral in Eq.~(\ref{integ}) more deeply, in particular, to
find where the integral in Eq.~(\ref{integ}) is saturated or what region of
integration is important, an approximation can be useful and worth mentioning.
The constant approximation for the function $f(t)$,
\begin{equation}
\label{f00}
f^{\rm appr}(t) = {\rm const} = f(0)= \frac{1}{\Delta}
\end{equation}
gives
\begin{equation}
I^{\rm appr}(\Delta) = f(0)\int_0^\infty F(t)dt
=f(0)\frac{m^2}{3}=\frac{m^2}{3\Delta} .
\end{equation}
This result represents the leading term of the series expansion of $I(\Delta)$
for small $m^2$. The series expansion of $I(\Delta)$ for small $m^2$ up to
terms of order $m^6$ is given by
\begin{equation}
I(\Delta) =\frac{1}{3}v+\left(\frac{19}{24}+\frac{1}{2}\ln v\right)v^2+
\left(\frac{77}{30}+2\ln v\right)v^3+\ldots
\end{equation}
with $v=m^2/\Delta$. This series converges nicely for small values of $v$.

The interpolation is only necessary for the light modes since the small $t$
behavior is nonperturbative. Heavy quarks can be treated in perturbation
theory as their masses are rather large~\cite{PDG,Kuhn:2001dm,mb}. The
contribution of $c$ and $b$ quarks to the MAMM reads~\cite{anom1}
\begin{equation}
a_\mu^{\rm had}({\rm LO;heavy}) = 71(18)\times 10^{-11}
\end{equation}
where we used $m_c=(1.6\pm 0.2)\GeV$ and $m_b=(4.8\pm 0.2)\GeV$. The
uncertainty mainly results from the uncertainty in the $c$ quark mass. Using
this result, the light mode contribution becomes
\begin{equation}
\label{normLO}
a_\mu^{\rm had}({\rm LO;light})=a_\mu^{\rm had}({\rm LO})
-a_\mu^{\rm had}({\rm LO;heavy}) = 6908(116)\times 10^{-11}.
\end{equation}
The values for the different cited values are shown in the third column of
Table~\ref{tab1}. Using this experimental result one finds a numerical value
for the IR parameter $\Delta$ of the interpolation function $f(t)$. Writing
$\Delta=4 m_{\rm eff}^2$ one obtains
\begin{equation}
\label{meffnum}
m_{\rm eff}=201.0\pm 1.9\MeV.
\end{equation}
The individual results are shown in the fourth column of Table~\ref{tab1}.
The function $r(s)$ of Eq.~(\ref{stepf}) is depicted in Fig.~\ref{fig5} for
$m_{\rm eff}=201\MeV$. This completes the quantitative description of the
interpolation function for the two-point correlator of the light modes which
can be used for the computation of the hadronic contributions at NLO. We cite
this interpolation as model~1 in the following.

\section{Hadronic contribution at next-to-leading order}
\begin{figure}\begin{center}
  \epsfig{figure=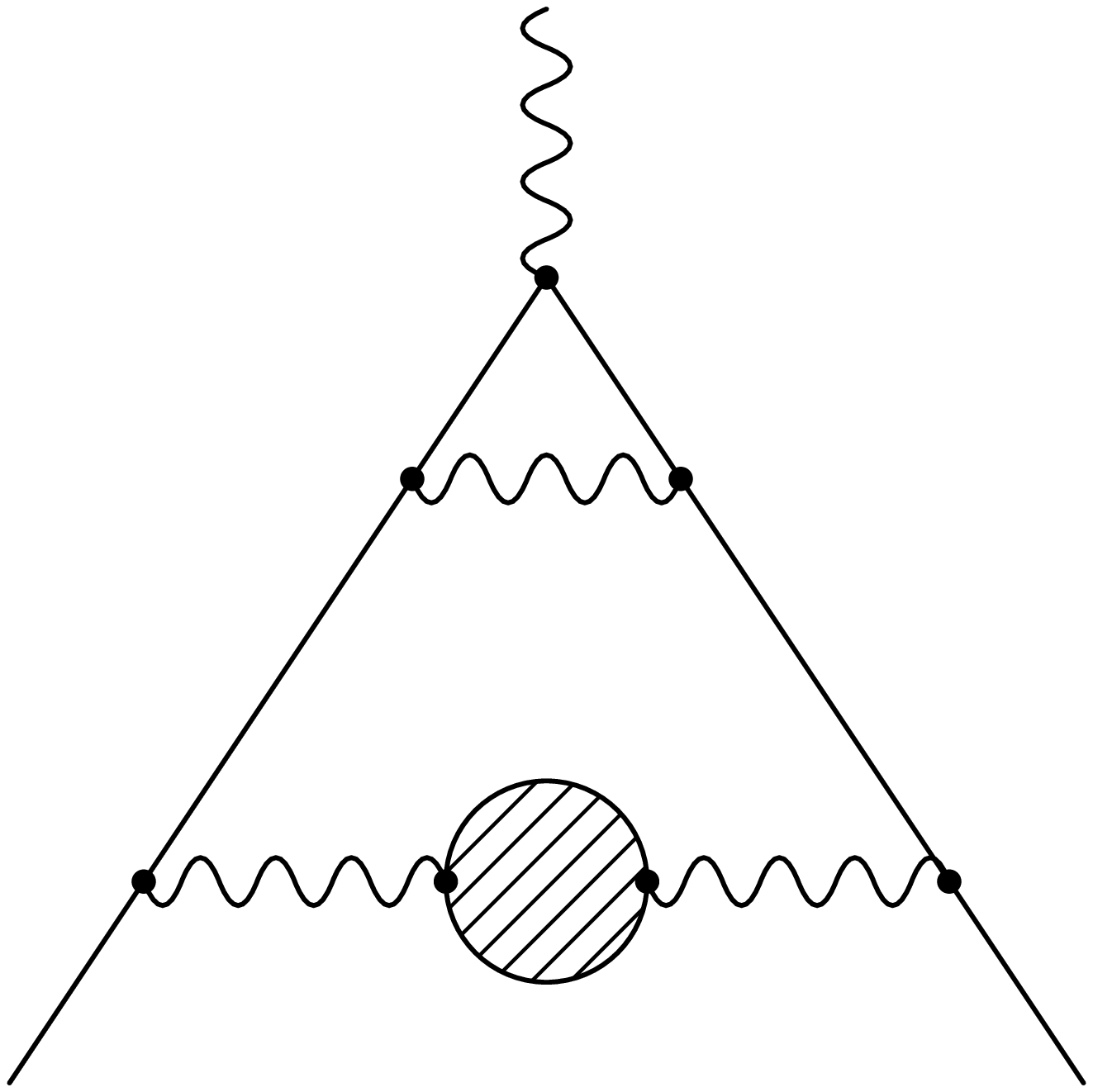, scale=0.2}\qquad
  \epsfig{figure=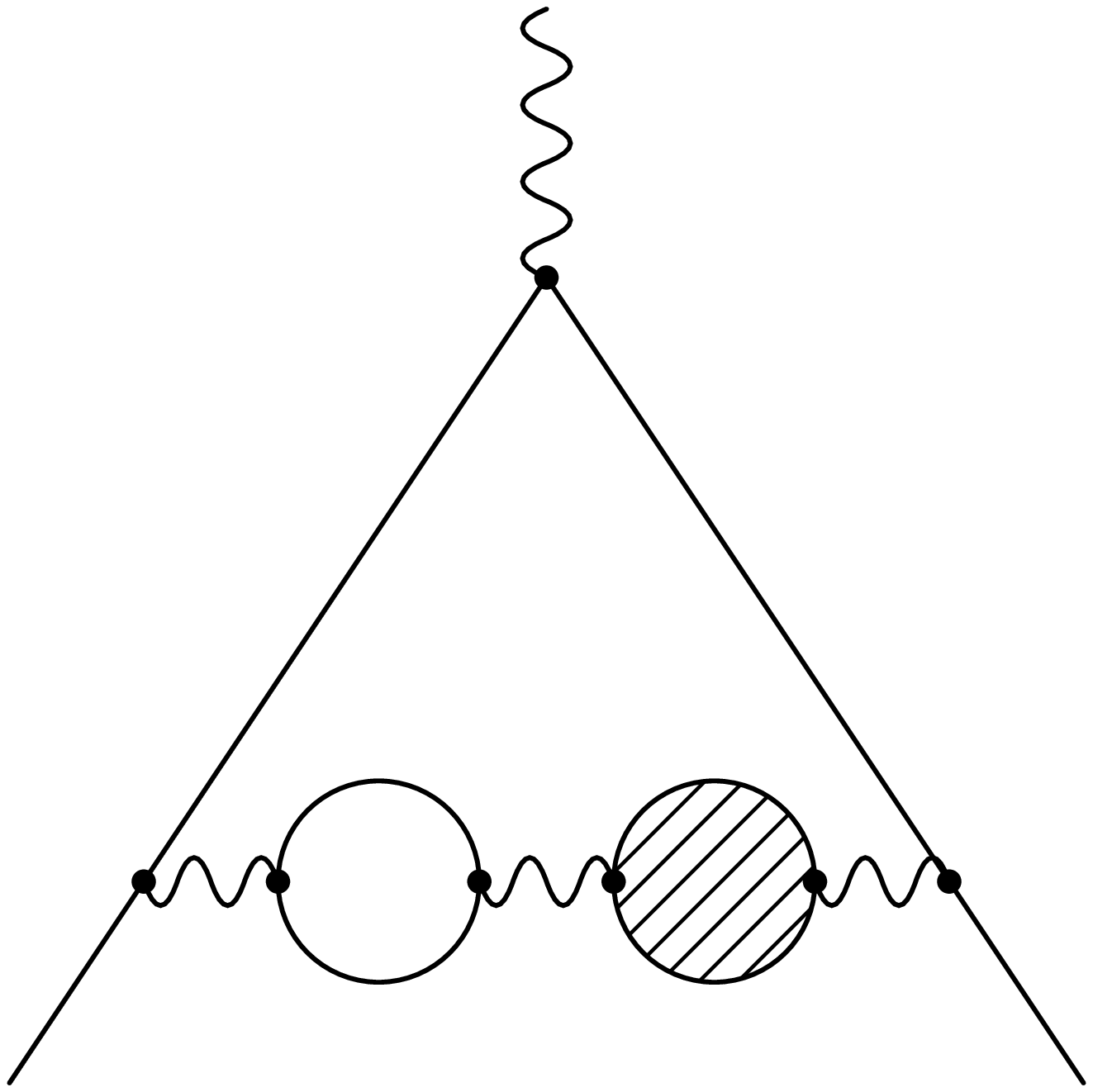, scale=0.2}
\caption{\label{fig3}NLO contributions to the MAMM involving the contribution
  of the hadronic two-point correlator (left) and a lepton-hadron type
  (so-called double bubble) diagram (right)}
\end{center}\end{figure}

The interpolation given by the function $f(t)$ for the two-point correlator
with the numerical value of the phenomenological parameter from
Eq.~(\ref{meffnum}) is now used at NLO. Two of the NLO diagrams involving
the hadronic two-point correlator are shown in Fig.~\ref{fig3}.
The NLO contribution is an integral of ${\rm Im}~\Pi^{\rm had}(s)$ with the
two-loop kernel $K^{(2)}(s)$,
\begin{equation}
\label{exp0vacNLO}
a_\mu^{\rm had}({\rm NLO})=4\pi\pfrac{\alpha}{\pi}^3\int_{0}^\infty
\frac{ds}{s}K^{(2)}(s){\rm Im}~\Pi^{\rm had}(s).
\end{equation}
The analytical expression for the kernel $K^{(2)}(s)$ is known~\cite{barbieri}.

Assuming that the IR scale $M_h$ of the hadronic spectrum
${\rm Im}~\Pi^{\rm had}(s)$ is larger than $m$ one can use an expansion of
$K^{(2)}(s)$ in $m^2/s$ under the integration sign in Eq.~(\ref{exp0vacNLO})
to generate an expansion in $m/M_h$ for the integral. The IR scale of the data
is set by the explicit cutoff at $\sqrt{s}=2m_\pi$. Then one generates an
expansion of $a_\mu^{\rm had}({\rm NLO})$ in $m^2/m_\pi^2$ using the data. The
data-based analysis for the NLO effects of the vacuum polarization
type~\cite{krause} gives
\begin{equation}
\label{vacpol2}
a_\mu^{\rm had}({\rm vac;NLO})=-101(6)\times 10^{-11}.
\end{equation}
In the proposed model the analysis is based on the explicit expression for the
hadronic two-point correlator given in Eq.~(\ref{expint}). In the model the
hadronic scale is given by $\sqrt{\Delta}=2m_{\rm eff}$ and the expansion of
the kernel results in the expansion of the integral in the variable
$m^2/\Delta$ or $m^2/m_{\rm eff}^2$. Note that convergence is not fast for the
hadronic scale given by $m_\pi$ or $m_{\rm eff}$. The integral in
Eq.~(\ref{exp0vacNLO}) can be rewritten in the Euclidean domain for
$\Pi^{\rm had}(q^2)$ in analogy to the LO treatment. For our purposes it
suffices to use the expansion in $m^2/s$. We present some contributions
separately for a comparison with the results from Ref.~\cite{krause}.

The vertex part of the kernel has an expansion~\cite{krause}
\[
K^{(2)}_{\rm ver}(s)=2\frac{m^2}{s}\left\{
\left(\frac{223}{54}-\frac{\pi^2}{3}-\frac{23}{36}\ln\pfrac{s}{m^2}\right)
\right.
\]
\[
+\frac{m^2}{s}\left(\frac{8785}{1152}-\frac{37\pi^2}{48}
-\frac{367}{216}\ln\pfrac{s}{m^2}+\frac{19}{144}\ln^2\pfrac{s}{m^2}\right)
\]
\begin{equation}
\label{expvert}
\left.
+\frac{m^4}{s^2}\left(\frac{13072841}{432000}-\frac{883\pi^2}{240}
-\frac{10079}{3600}\ln\pfrac{s}{m^2}+\frac{141}{80}\ln^2\pfrac{s}{m^2}\right)
+\ldots\right\}.
\end{equation}
Generally, the terms of the expansion contain powers and logarithms of the
variable $m^2/s$. For pure powers one can use a generating integral
representation with a polynomial $P(x)$
\begin{equation}
\label{ppow}
m^2 \int_0^1\frac{dx P(x)}{m^2 x+s}
=\frac{m^2}{s}\sum_n a_n \pfrac{m^2}{s}^n,
\qquad
a_n=\int_0^1 dx P(x)(-x)^n.
\end{equation}
A given polynomial $P(x)$ restores the pure power expansion of
Eq.~(\ref{expvert}). For the logarithmic part the generating integral
representation can be chosen with a polynomial $G(x)$ of the form
\begin{equation}
\label{gg1g2}
m^2 \int_0^1\frac{dx G(x)}{s x+m^2}=G_1(m^2/s)+G_2(m^2/s)\ln\pfrac{s}{m^2}.
\end{equation}
The polynomial $G(x)$ generates polynomials $G_1(x)$, $G_2(x)$ through
Eq.~(\ref{gg1g2}). The mixture of pure powers due to the polynomial $G_1(x)$
leads to a redefinition of the polynomial $P(x)$ in Eq.~(\ref{ppow}). Using
Eqs.~(\ref{exp0vacNLO},\ref{ppow}) one finds the expression for the pure power
part of the expansion
\begin{equation}
\label{deri}
\frac{1}{\pi}\int_0^\infty \frac{ds}{s}K^{(2)}(s)|_{\rm power}
{\rm Im}~\Pi^{\rm had}(s)=\int_0^1 \frac{dx}{x}P(x)[-\Pi^{\rm had}(-m^2 x)]
\end{equation}
which reduces to derivatives of $\Pi^{\rm had}(t)$ at the origin and gives the
analytic part of the expansion in $m/M_h$. For the logarithmic part one finds
the representation
\begin{equation}
\label{intlog}
\frac{1}{\pi}\int_0^\infty \frac{ds}{s}K^{(2)}(s)|_{\rm power\&log}
{\rm Im}~\Pi^{\rm had}(s)=
\int_0^1 [-\Pi^{\rm had}(-m^2/x)] G(x)dx
\end{equation}
which is sensitive to the entire Euclidean domain and gives the nonanalytic
part of the expansion containing $\ln(m/M_h)$. This procedure can be performed
up to any finite order in $m^2$ and the whole calculation can be organized in
a way such that only Euclidean values of momenta are necessary for
$\Pi^{\rm had}(q^2)$. Derivatives of the function $\Pi^{\rm had}(t)$ at the
origin emerging from Eq.~(\ref{deri}) depend on the hadronic scale $M_h$ while
the muon mass enters polynomially. The integral in Eq.~(\ref{intlog}) depends
on both the hadronic scale and the muon mass. The behavior of
$\Pi^{\rm had}(q^2)$ in the Euclidean domain is smooth and perturbative at
large momenta. The region near the origin $q^2=0$ is a nonperturbative one.
Thus, the basic objects that emerge in the analysis are derivatives of the
function $\Pi^{\rm had}(t)$ at the origin and integrals of the form
\begin{equation}
\int_{m^2}^\infty  \frac{dt}{t^n}\Pi^{\rm had}(-t)
\ln^p\pfrac{t}{m^2}.
\end{equation}
One can use this technique to avoid any reference to the physical region.

In model~1 given by Eqs.~(\ref{expint},\ref{modstep}) the explicit
interpolation function $p(t)$ is given in the whole complex $t$-plane.
Therefore, it makes no difference how one computes the necessary integrals
either in the Minkowskian or in the Euclidean domain. At the formal
mathematical level the calculation with an explicit function can be performed
in the spectral representation alone. The Euclidean approach will lead to the
same formal results. For the calculation of the basic objects emerging in the
spectral representation
\begin{equation}
\label{sperepmom}
{\cal M}_{n,p}(\Delta)=\Delta^n
\int_\Delta^\infty \frac{ds}{s^{n+1}}\ln^p \pfrac{s}{m^2}
\end{equation}
the recurrence relation
\begin{equation}
{\cal M}_{n,p}(\Delta)=\frac{1}{n}\ln^p(\Delta/m^2)
+\frac{p}{n}{\cal M}_{n,p-1}(\Delta)
\end{equation}
can be used to decrease a power of the logarithm in the integrand of
Eq.~(\ref{sperepmom}). The results for the first two powers of the logarithm
are
\begin{equation}
{\cal M}_{n,0}(\Delta)=\frac1n,\qquad
{\cal M}_{n,1}(\Delta)=\frac1n\ln(\Delta/m^2)+\frac1{n^2}.
\end{equation}
Writing the analytical expression for the NLO vertex contribution to the MAMM
in the form
\begin{equation}
\label{stepex}
a_\mu^{\rm mod}({\rm ver;NLO;analyt})=\pfrac{\alpha}{\pi}^3
V(m^2/\Delta)
\end{equation}
(the index ``mod'' stands for model~1, ``ver'' represents the vertex
contribution in Fig.~\ref{fig3} left) one finds
\begin{equation}
\label{stepvexp}
V(v)=\frac{v}{9}\left(\frac{377}{18}-2\pi^2 + \frac{23}{6}\ln v\right)
+\frac{v^2}{9}
\left(\frac{23647}{1152}-\frac{37\pi^2}{16} + \frac{677}{144}\ln v
+ \frac{19}{48}\ln^2 v\right)+o(v^2)
\end{equation}
where $o(v^2)$ is any function that satisfies $\lim_{v\to 0}o(v^2)/v^2=0$. For
brevity we have explicitly presented only two terms of the expansion of the
function $V(m^2/\Delta)$ at small $m^2/\Delta$ resulting from the
corresponding expansion of the kernel in Eq.~(\ref{expvert}). The result of
evaluating the $u$, $d$, $s$ light mode contribution obtained from
Eqs.~(\ref{stepex},\ref{stepvexp}), together with the appropriate QCD group
factor $N_c(e_u^2+e_d^2+e_s^2)=2$ reads
\begin{equation}
\label{apprstepex}
a_\mu^{\rm mod}({\rm ver;NLO;light;analyt})=-191\times 10^{-11}.
\end{equation}
The numerical integration of the kernel given in Eq.~(\ref{expvert}) results
in the value
\begin{equation}
\label{intdir3}
a_\mu^{\rm mod}({\rm ver;NLO;light})=-190(2)\times 10^{-11},
\end{equation}
to be compared with the corresponding results from the model of
Ref.~\cite{anom1} which is based on the use of free massive fermions for 
computation of the interpolation function,
\begin{equation}
a_\mu^{\rm mod}({\rm ver;NLO;light})=-188\times 10^{-11}.
\end{equation}
The total contribution of the vertex type including the heavy quarks reads
\begin{equation}
a_\mu^{\rm mod}({\rm ver;NLO})=-194(3)\times 10^{-11},
\end{equation}
to be compared with the one obtained in the data-based approach~\cite{krause}
\begin{equation}
a_\mu^{\rm Ref.~\cite{krause}}({\rm ver;NLO})=-211(5)\times 10^{-11}.
\end{equation}
Note that the leading order contribution used in Ref.~\cite{krause} is
different from the value in Eq.~(\ref{normLO}). Therefore the direct
comparison should be made with some caution. In fact, the value used for the
LO contribution in Ref.~\cite{krause} is larger than the value in
Eq.~(\ref{normLO}) which would result in a smaller value of $m_{\rm eff}$ and
consequently the larger value of the NLO vertex correction in the model
calculation given in Eqs.~(\ref{apprstepex},\ref{intdir3}). Nevertheless, the
difference is within the error bars.

For a mixed contribution of the lepton-hadron type (so-called double bubble
(``db'') diagram) shown on the right hand side in Fig.~\ref{fig3} we find
\begin{equation}
\label{elpol}
a_\mu^{\rm mod}({\rm db;NLO;lept+had})=106(2)\times 10^{-11}
\end{equation}
while the data-based estimate reads
\begin{equation}
a_\mu^{\rm Ref.~\cite{krause}}({\rm bd;NLO;lept+had})=107(2)\times 10^{-11}.
\end{equation}
Thus, the results obtained with the interpolation model given in
Eqs.~(\ref{expint},\ref{modstep}) reproduce those obtained in the data-based
approach.

The result for the total NLO hadronic contribution of the vacuum polarization
type (two-point correlator only) is
\begin{equation}
a_\mu^{\rm mod}({\rm vac;NLO})=-86.5(0.7)\times 10^{-11}.
\end{equation}
which has to be compared with the result shown in Eq.~(\ref{vacpol2}). All
contributions, including the leading order contributions which reproduce the
input of the determination of $m_{\rm eff}$ are listed in the second column
of Table~\ref{tab2}.

\begin{table}[t]\begin{center}
\begin{tabular}{|r||c|c|c|}\hline
  contribution&model~1&model~2&model~3\\\hline\hline
  $a_\mu^{\rm mod}({\rm LO;light})$&$6908(114)\times 10^{-11}$
    &$6908(114)\times 10^{-11}$&$6578\times 10^{-11}$\\\hline
  $a_\mu^{\rm mod}({\rm LO;heavy})$
    &\multicolumn{3}{|c|}{$71(18)\times 10^{-11}$}\\\hline\hline
  $a_\mu^{\rm mod}({\rm ver;NLO;light})$&$-189.5(2.4)\times 10^{-11}$
    &$-188.2(2.4)\times 10^{-11}$&$-183.2\times 10^{-11}$\\\hline
  $a_\mu^{\rm mod}({\rm ver;NLO;heavy})$
    &\multicolumn{3}{|c|}{$-4.3(0.9)\times 10^{-11}$}\\\hline
  $a_\mu^{\rm mod}({\rm db;NLO;light})$&$105.0(1.7)\times 10^{-11}$
    &$104.9(1.7)\times 10^{-11}$&$100.0\times 10^{-11}$\\\hline
  $a_\mu^{\rm mod}({\rm db;NLO;heavy})$
    &\multicolumn{3}{|c|}{$1.1(0.3)\times 10^{-11}$}\\\hline\hline
  $a_\mu^{\rm mod}({\rm vac;NLO})$&$-87.7(0.7)\times 10^{-11}$
    &$-86.5(0.7)\times 10^{-11}$&$-86.4\times 10^{-11}$\\\hline
\end{tabular}\end{center}
\caption{\label{tab2}The different vacuum polarization type LO and NLO
  contributions for model~1 (second column), model~2 (third column), and
  model~3 (fourth column). The values for the effective masses for model~1
  and model~2 are taken as $m_{\rm eff}=(201.0\pm 1.9)\MeV$ and
  $m_q=(179.4\pm0.6)\MeV$ respectively, the cited uncertainties are a
  consequence of the uncertainties of these effective masses and of the masses
  of the heavy quarks. The error estimates due to these two sources can be
  added quadratically.}
\end{table}

\begin{figure}
\epsfig{figure=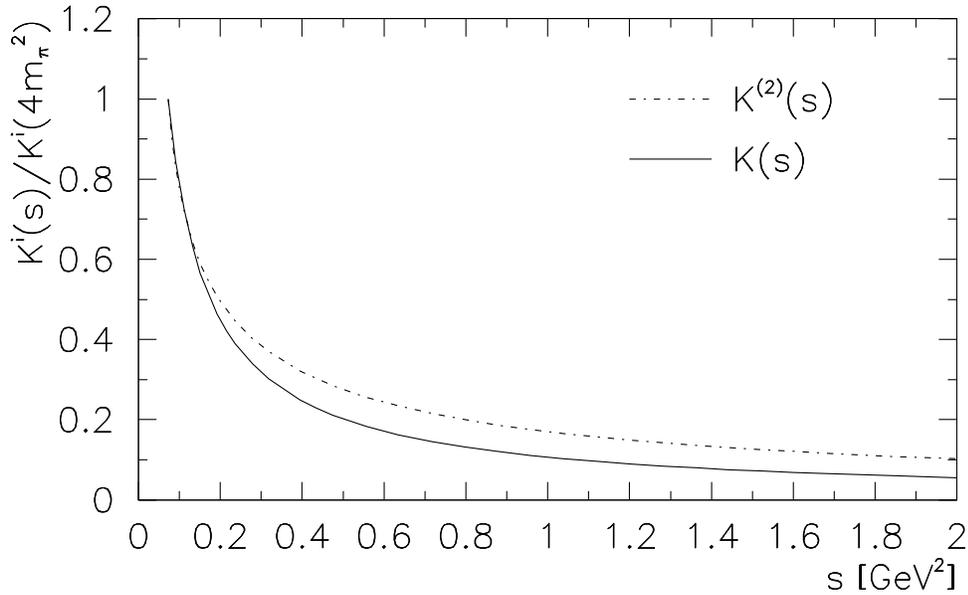, scale=0.8}
\caption{\label{fig4}The LO and NLO kernels $K(s)$ and $K^{(2)}(s)$.}
\end{figure}

The model given in Eqs.~(\ref{expint},\ref{modstep}) reproduces rather
accurately the results for the NLO hadronic contributions found in the
data-based analysis for the graphs related to vacuum polarization. This has
been expected as these results are obtained by the integration of the two-point
function with the NLO kernel $K^{(2)}(s)$ which has a structure very close to
that of the leading order kernel $K(s)$. These two functions are shown in
Fig.~\ref{fig4}.

\section{Alternatives for the model spectral function}
\begin{figure}
\epsfig{figure=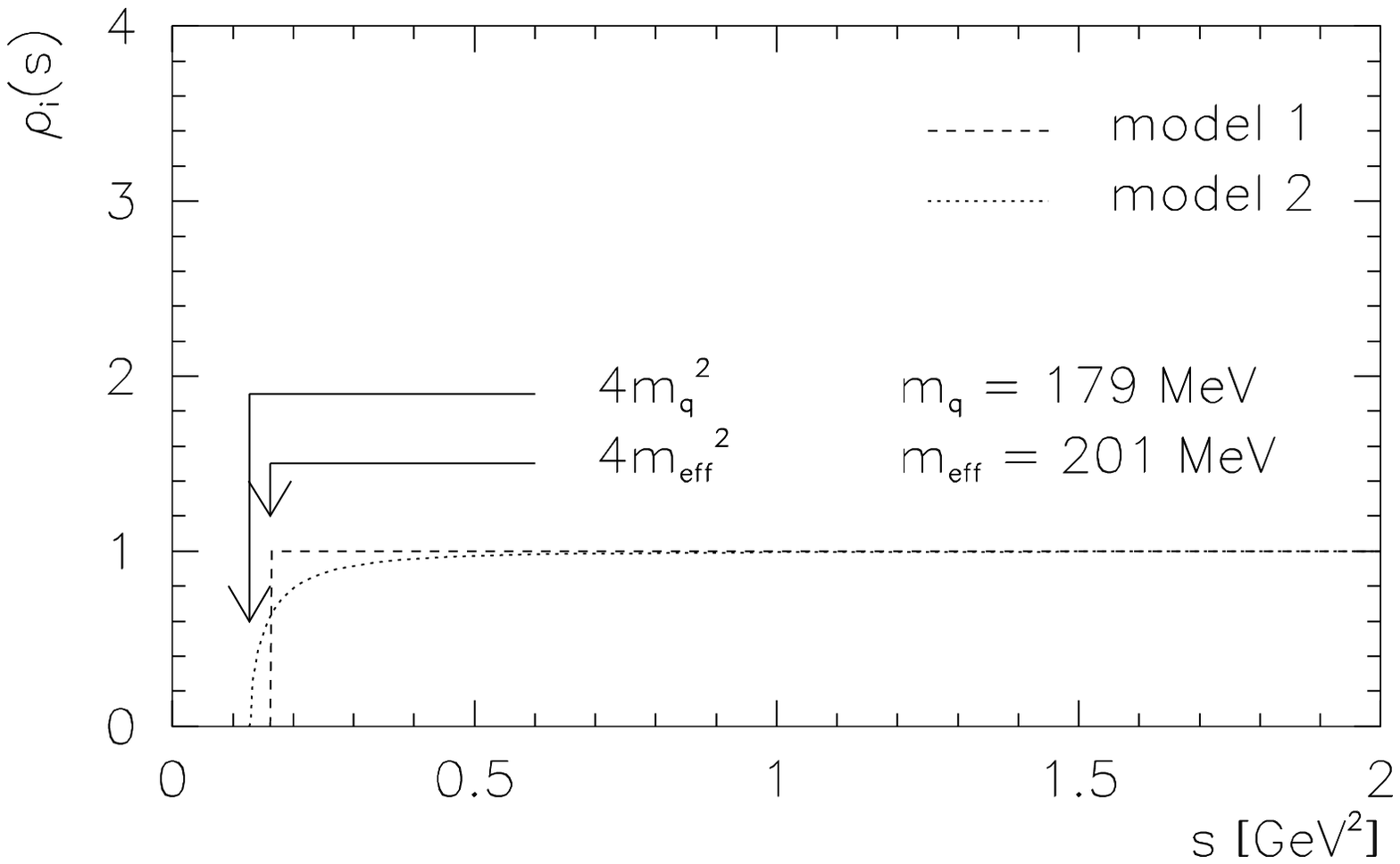, scale=0.8}\\
\epsfig{figure=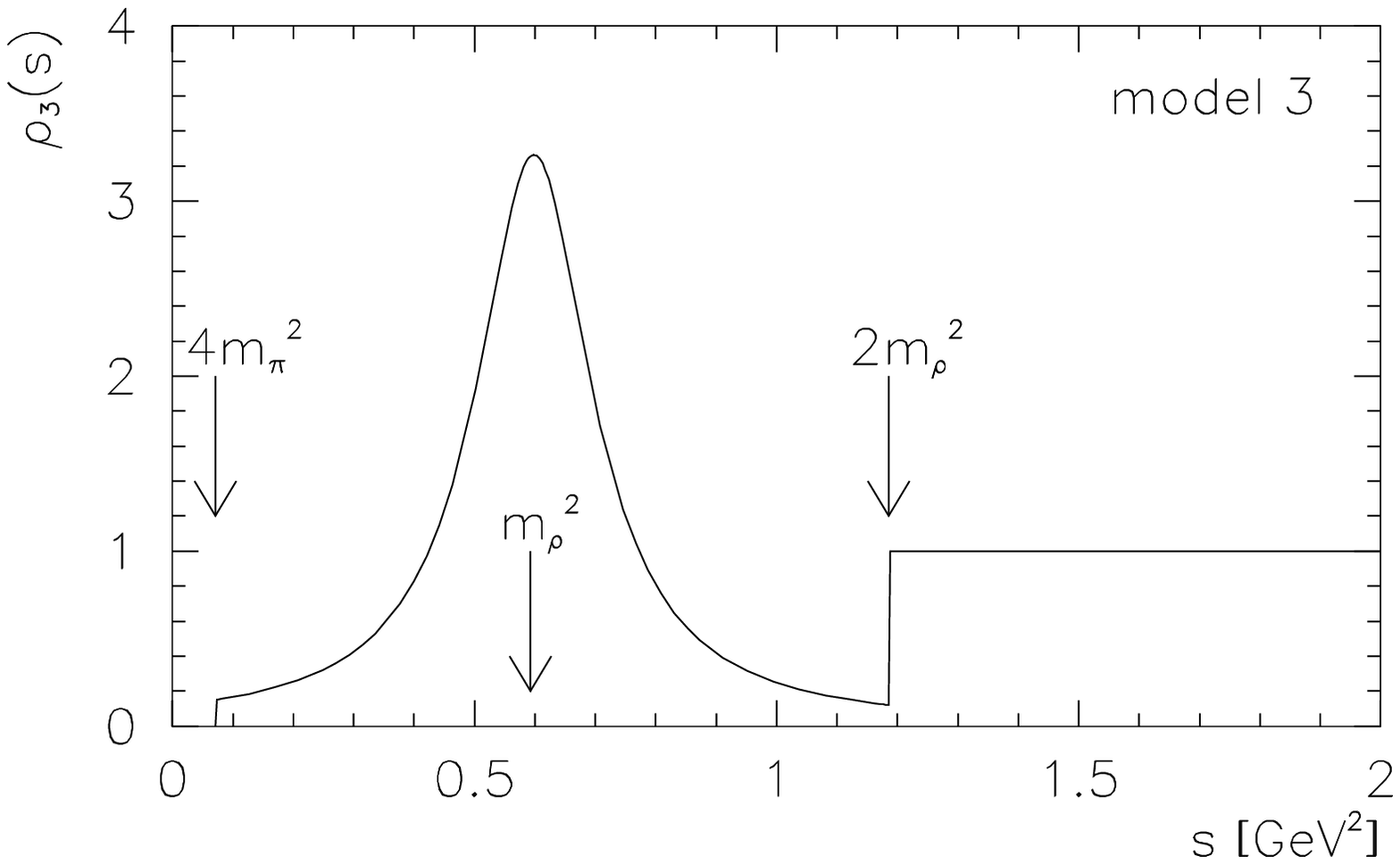, scale=0.8}
\caption{\label{fig5}$s$-dependence of the spectral functions $\rho_1(s)=r(s)$
  of model~1 in Eq.~(\ref{stepf}) and $\rho_2(s)=3\rho_q(s)$ of model~2 in
  Eq.~(\ref{fermspect}) (upper diagram), as compared to the spectral function
  $\rho_3(s)=\rho^{\rm had}(s)$ for model~3 in
  Eqs.~(\ref{sphadmod},\ref{bwspect}). We use $m_{\rm eff}=201\MeV$,
  $m_q=179\MeV$, and the central values $m_\rho=769.9\MeV$ and
  $\Gamma_\rho=150.2\MeV$~\cite{PDG}.}
\end{figure}

The interpolation function for the two-point correlator of hadronic EM
currents in Eq.~(\ref{expint}) is very simple. One can use more sophisticated
interpolations. A formal criterion for the choice of the interpolation is its
consistency with general principles of quantum field theory (analyticity and
unitarity in this case). A practical criterion is its simplicity such that
analytical calculations become technically feasible. One can turn to free
field models in a search for mathematical functions that can be used in the
interpolation procedure. For instance, the scalar or fermionic polarization
functions with masses as free parameters can be taken as suitable candidates.
The fermionic interpolation function was considered in detail in
Ref.~\cite{anom1}. It is given by the expression
\[
\pi(t,m_q)=\left(\frac{1}{3z}-1\right)\varphi(z)-\frac{1}{9},
\]
\begin{equation}
\label{lasttech}
\varphi(z)=\frac{1}{\sqrt{z}}\artanh(\sqrt{z})-1,\quad
z=\frac{t}{4m_q^2+t}.
\end{equation}
The discontinuity across the cut $(4m_q^2,\infty)$ at $t=-s-i0$ is given by
the fermionic spectral density of the form 
\begin{equation}
\label{fermspect}
\rho_q(s)=\frac{1}{3}
\sqrt{1-\frac{4 m_q^2}{s}}\left(1+\frac{2 m_q^2}{s}\right).
\end{equation}
A pictorial representation of $\rho_q(s)$ is shown in Fig.~\ref{fig5}.
The two functions $f(t,m_{\rm eff})/3$ and $-d\pi(t,m_q)/dt$ coincide within
1\% accuracy in the interval $t=(0,m_q^2)$ if the effective parameters are
related through $m_{\rm eff}/m_q=\sqrt{5}/2\approx 1.12$. This is expected
from duality considerations based on the shape of the discontinuity across the
cut given in Eqs.~(\ref{stepf},\ref{fermspect}). Since the integral in
Eq.~(\ref{integ}) is saturated at the scale of the order of the muon mass
$\sqrt{t}\sim m_\mu$ and $m_\mu<m_{\rm eff}$, the closeness of the integrated
interpolation functions implies a rather accurate equality of the resulting
integrals. The generalization to interpolations based on scalar field theories
is straightforward. The interpolation given by the spectral density in
Eq.~(\ref{fermspect}) will be referred to as model~2 in the following.

At this stage it seems that hadrons have completely disappeared from the
analysis and some artificial functions are rather arbitrarily used to compute
the relevant integrals. The link to physics is that for such kind of inclusive
observables as the MAMM which are sensitive to a contribution of many hadrons
the analysis can completely be done in the Euclidean domain with the only IR
sensitive contribution coming from the region near the origin. In the
Euclidean domain the contribution of all hadrons to the MAMM is smeared to the
extent which is determined by the distance between the integration region and
the nearest physical singularity. The hadronic singularity is taken from
experiment as a two-pion cut with the kinematical constraint $s>4m_\pi^2$. The
integration, in fact, is sensitive to scales $\sqrt{t}\sim 2m_\mu$. Therefore,
a correctly normalized function (as for duality at large energies) with the
appropriate IR behavior in the interval $0<t<4m_\mu^2$ reproduces the data with
reasonable accuracy. The IR behavior is mainly determined by the experimental
scale $4m_\pi^2$.

To illustrate this statement let us consider a more realistic interpolation
for the vacuum polarization function in the Euclidean domain from the point of
view of experimental hadron physics than those described so far. The model
hadron spectrum for light modes can be chosen in the following simple form
\begin{equation}
\label{sphadmod}
\rho^{\rm had}(s)=2m_\rho^2\delta(s-m_\rho^2)+\theta(s-2 m_\rho^2).
\end{equation}
This is a one-scale no-parameter model that satisfies the duality constraints
from the operator product expansion~\cite{Shif,Kra1983,locdual}. The hadronic
scale of the model is given by the $\rho$-meson mass $m_\rho$ which is
eventually fixed from experiment~\cite{PDG}. We neglect small violations of
flavor symmetry for the $u,d,s$ light modes and consider them to be
degenerate. The spectrum from Eq.~(\ref{sphadmod}) gives an interpolation
function of the form
\begin{equation}
f^{\rm had}(t)=\frac{2m_\rho^2}{(t+m_\rho^2)^2}
+\frac{1}{t+2m_\rho^2}.
\end{equation}
The value of the interpolation function $f^{\rm had}(t)$ at the origin $t=0$
reads
\begin{equation}
\label{infnarr}
f^{\rm had}(0)=\frac{5}{2m_\rho^2}.
\end{equation}
It should be compared with the value from Eq.~(\ref{f00}). For
$m_\rho=769.9\MeV$ one finds $f^{\rm had}(0)=4.22~{\rm GeV^{-2}}$ while
the data give $f(0)=1/\Delta=6.25~{\rm GeV^{-2}}$. This is fairly reasonable
given the simplicity of the model but is not accurate enough. Note that there
is no single free parameter in the model given in Eq.~(\ref{sphadmod}) and the
shape of the spectrum is fixed from the duality constraint at large energies.
The weight function $F(t)$ from Eqs.~(\ref{ffuncparts},\ref{expliffun})
determining the MAMM integral can resolve the behavior of the interpolation
function $f^{\rm had}(t)$ for the hadron correlator at the scales of order
$m_\mu=105.66\MeV$. However, the approximation of infinitely narrow resonance
in Eq.~(\ref{sphadmod}) is too rough for computing such an integral. Thus, the
low energy behavior of the spectrum is not precise enough which leads to an
insufficient accuracy of the interpolation function at small $t$. Formally
this is seen in the absence of the scale $4m_\pi^2$ which is known to be
important for the evaluation of the integral in the data-based analysis.
Therefore the spectrum from Eq.~(\ref{sphadmod}) should be corrected for this
particular application. A dominant role in the data-based analysis is played
by two-pion states. The $\rho$ meson is a resonance in the two-pion system
therefore its contribution effectively takes into account the pion singularity
as well. However, the zero width approximation is not good enough for
computing the particular integral in Eq.~(\ref{ffuncparts}). A natural
modification of the spectrum is to introduce a finite width for the $\rho$
meson. This is achieved by replacing the function $\delta(s-m_\rho^2)$ by the
Breit-Wigner function for the resonance part of the spectrum in
Eq.~(\ref{sphadmod})
\begin{eqnarray}
\label{bwspect}
\rho^{\rm had}_{\rm R}(s)&=&\frac{2m_\rho^2}{\pi}
\frac{\Gamma_\rho m_\rho}{(s-m_\rho^2+\Gamma_\rho^2/4)^2
  +\Gamma_\rho^2 m_\rho^2},\nonumber\\
\rho^{\rm had}_{\Gamma}(s)&=&\theta(s-4m_\pi^2)\theta(2m_\rho^2-s)
\rho^{\rm had}_{\rm R}(s) +\theta(s-2m_\rho^2).
\end{eqnarray}
The interpolation function based on this spectrum will be called model~3 in
the following. Fig.~\ref{fig5} shows the $s$-dependence of
$\rho^{\rm had}_\Gamma(s)$. The expression for the resonance part of the
spectrum reduces to $2m_\rho^2\delta(s-m_\rho^2)$ in the limit
$\Gamma_\rho\to 0$. Using the Breit-Wigner form of the spectrum for the region
$4m_\pi^2<s<2 m_\rho^2$ one finds the contribution of the resonance to the
interpolation function in the Euclidean domain
\begin{equation}
\label{bwinter}
f^{\rm had}_{\rm R}(t)=\int\limits_{4 m_\pi^2}^{2m_\rho^2}
\frac{\rho^{\rm had}_{\rm R}(s)ds}{(s+t)^2}.
\end{equation}
The interpolation function in the Euclidean domain for the spectrum with
nonzero width reads
\begin{equation}
\label{bwintertot}
f^{\rm had}_{\Gamma}(t)=f^{\rm had}_{\rm R}(t)+\frac{1}{t+2 m_\rho^2}.
\end{equation}
Computing the value of the interpolation function at the origin for
$\Gamma_\rho=150.2\MeV$~\cite{PDG} one finds
\begin{equation}
\label{widthg}
f^{\rm had}_\Gamma(0)=f^{\rm had}_{\rm R}(0)+\frac{1}{2 m_\rho^2}
=5.15+0.84=6.0~{\rm GeV^{-2}}
\end{equation}
instead of the result~(\ref{infnarr}) obtained in the infinitely narrow
resonance approximation. The number from Eq.~(\ref{widthg}) differs from the
data-based estimate $f(0)=1/\Delta=6.25~{\rm GeV^{-2}}$ by 4\% only. Note that
no free parameters have been used so far for the description of the hadron
spectrum. The integrals entering the MAMM as in Eq.~(\ref{integ}) are also
rather close. One finds
\begin{equation}
\label{dataint}
I^{\rm data}=0.0194
\end{equation}
to be compared with
\begin{equation}
I^{\rm had}_\Gamma=0.0155+0.0030=0.0185
\end{equation}
where the first term in the sum is given by the resonance and the second by
the continuum contribution. Thus, the simple and parameter-free model from
Eqs.~(\ref{bwspect},\ref{bwinter}) already gives a reasonable precision of
about 5\% for the LO hadronic contribution to the MAMM. In this sense it
successfully incorporates experimental information necessary for the MAMM
computation.

\begin{figure}
\epsfig{figure=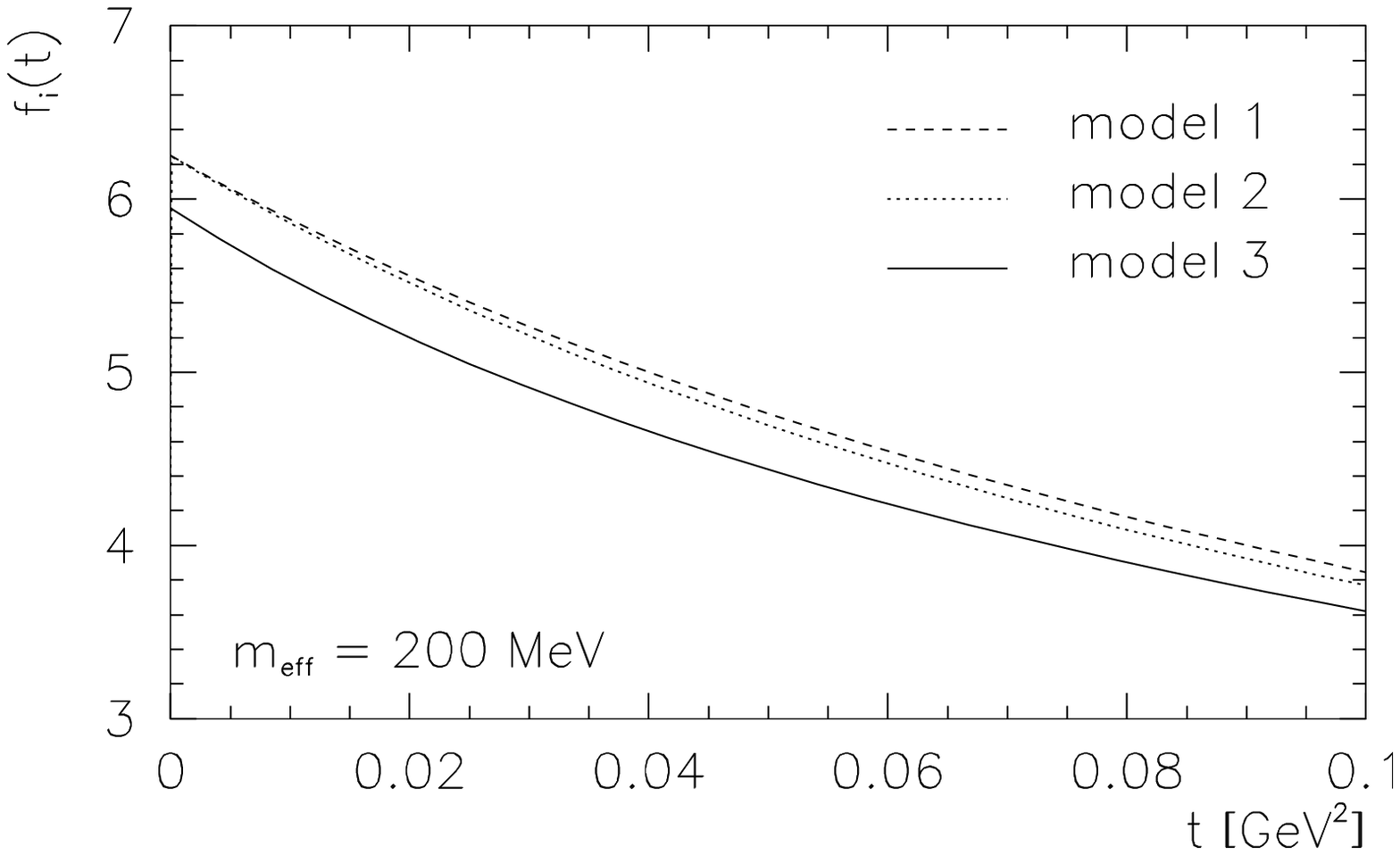, scale=0.8}\\
\epsfig{figure=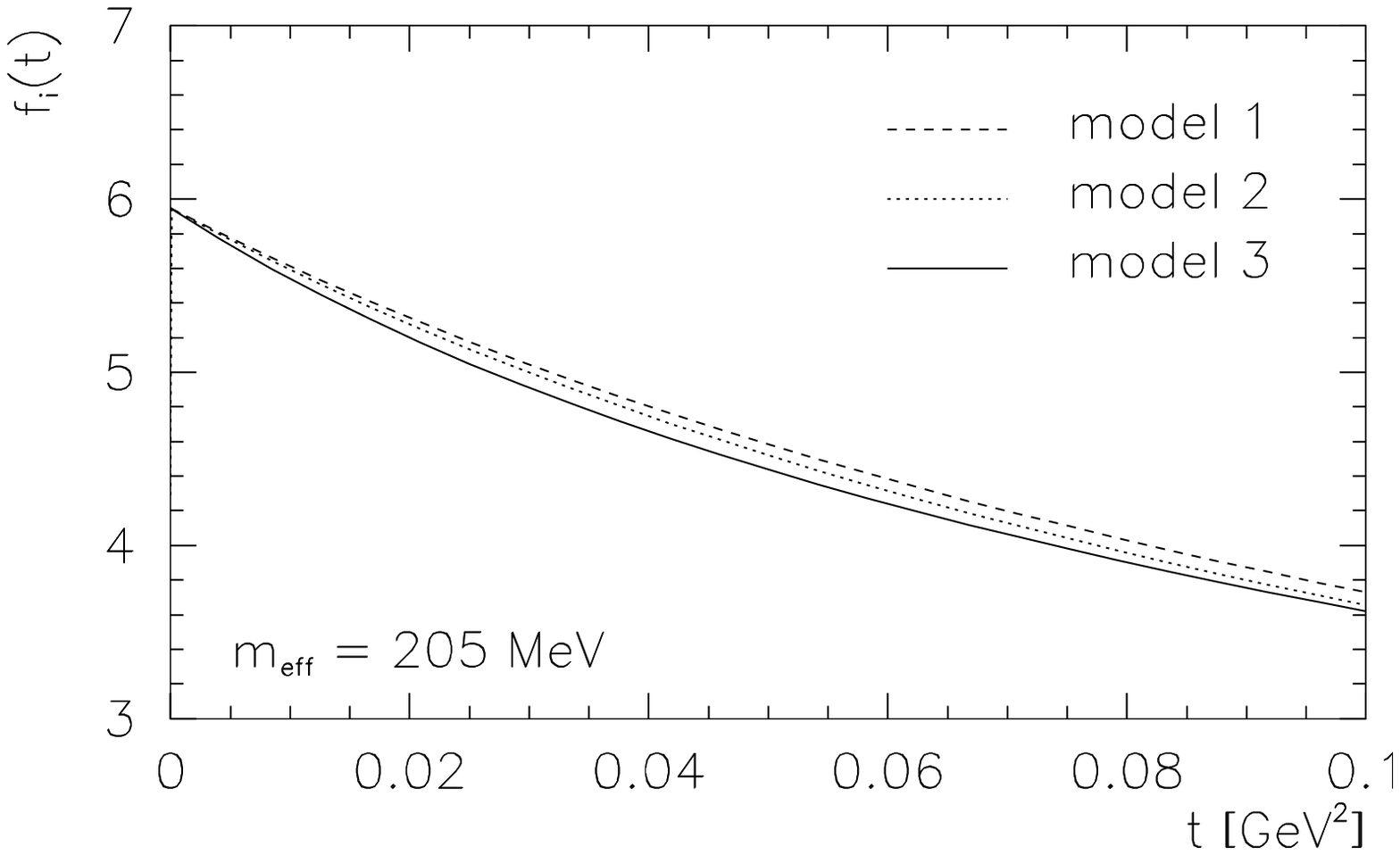, scale=0.8}
\caption{\label{fig6}The functions $f_i(t)$ for the three different models
  where $f_1(t)=f(t)$ is given by Eq.~(\ref{expint}), $f_2(t)=-3d\pi(t)/dt$ is
  given by Eq.~(\ref{lasttech}) and $f_3(t)=f_\Gamma^{\rm had}(t)$ is given by
  Eq.~(\ref{bwintertot}). We take $\Delta=4m_{\rm eff}^2$ where
  $m_{\rm eff}=200\MeV$ is used for the upper diagram and
  $m_{\rm eff}=205\MeV$ for the lower diagram. The parameter $m_q$ used for
  $\pi(t)$ is connected to $m_{\rm eff}$ by $m_q=2m_{\rm eff}/\sqrt 5$. The
  values $m_\rho=769.9\MeV$ and $\Gamma_\rho=150.2\MeV$ used in
  $f_R^{\rm had}$ are taken from Ref.~\cite{PDG}.}
\end{figure}

The spectral functions taken for model~1 from Eq.~(\ref{stepf}), for model~2
from Eq.~(\ref{fermspect}), and for model~3 from Eq.~(\ref{bwspect}) are shown
in Fig.~\ref{fig5} in order to allow one to compare these models. Neither
model~1 nor model~2 has a discontinuity across the positive semiaxis of the
$s$-plane resembling the experimental spectrum. However, both models result in
integrals over the spectrum for the respective kernels which are very close to
the result obtained in the experimentally inspired model~3 and, eventually, to
the data represented in Eq.~(\ref{dataint}). From the purely mathematical
point of view this is related to the fact that the procedure of analytic
continuation is an incorrectly posed problem: small variations of functions in
the Euclidean domain can produce big variations on the cut. The $t$-dependence
of the Euclidean representation by the functions $f(t)$, $-3d\pi(t)/dt$, and
$f^{\rm had}_\Gamma(t)$ is shown in Fig.~\ref{fig6}. A phenomenological
interpretation of the situation consists in a duality between hadrons and free
light fermions with QCD quantum numbers as for the particular application
related to the MAMM computation. Of course, the main objective of the
calculation of the hadronic contribution at the leading order from experiment
is to reach a high precision. The use of direct data seems to be superior
to a parameterization of the spectrum from indirect observations. However, as
soon as the integral over the data is computed a smooth interpolation function
of a simple form can be introduced in the Euclidean domain to be used in
higher order calculations. Because this interpolation function is explicit and
complies with the general properties of analyticity and unitarity one can find
its discontinuity across the positive semiaxis and perform further
calculations in the spectral representation as well. The analysis of NLO
contributions along these lines shows that the data-based results are
accurately reproduced~\cite{anom1}.

\section{Discussion of the light-by-light contribution}
\begin{figure}
\centerline{\epsfig{figure=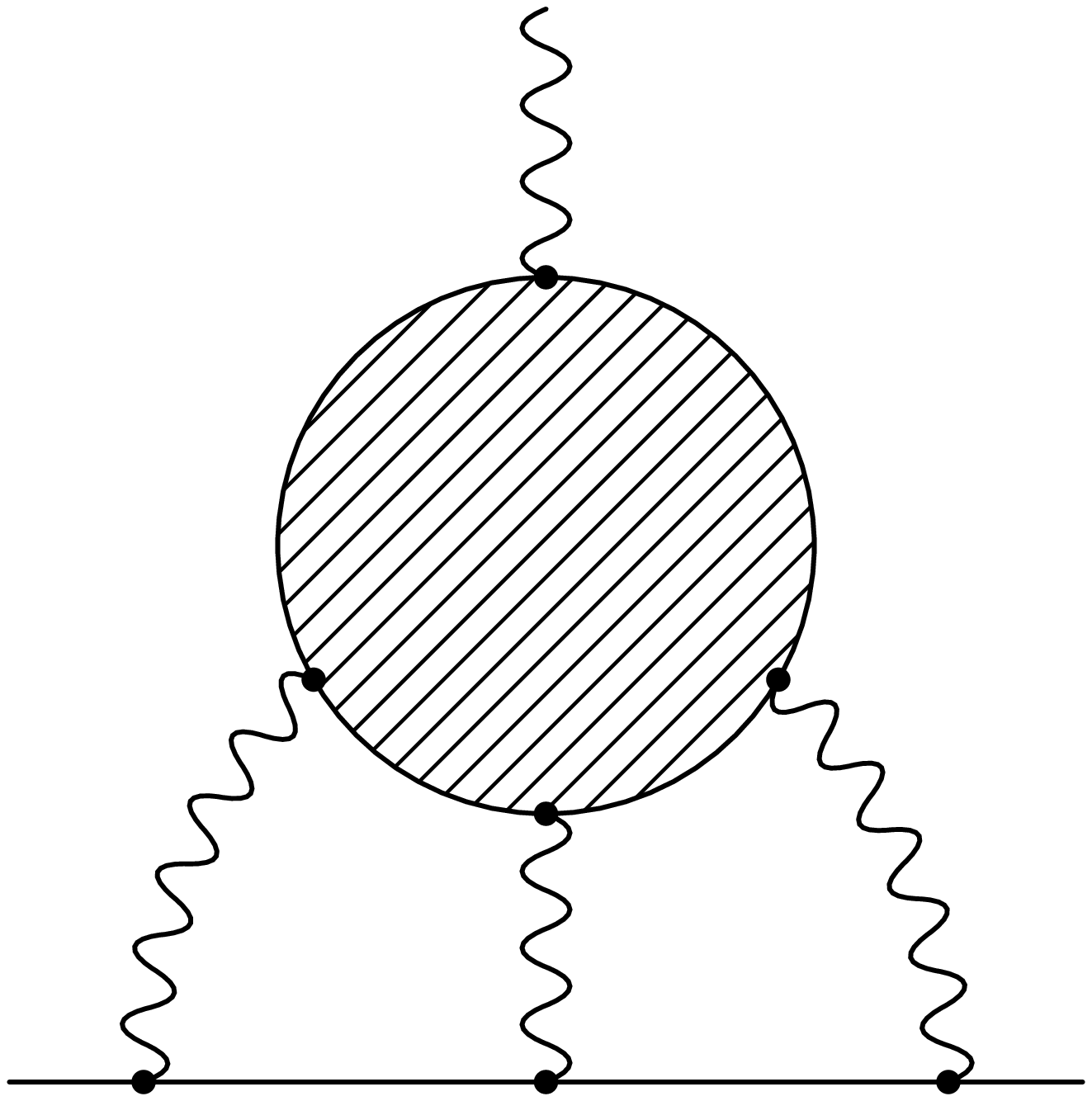, scale=0.2}\qquad
  \epsfig{figure=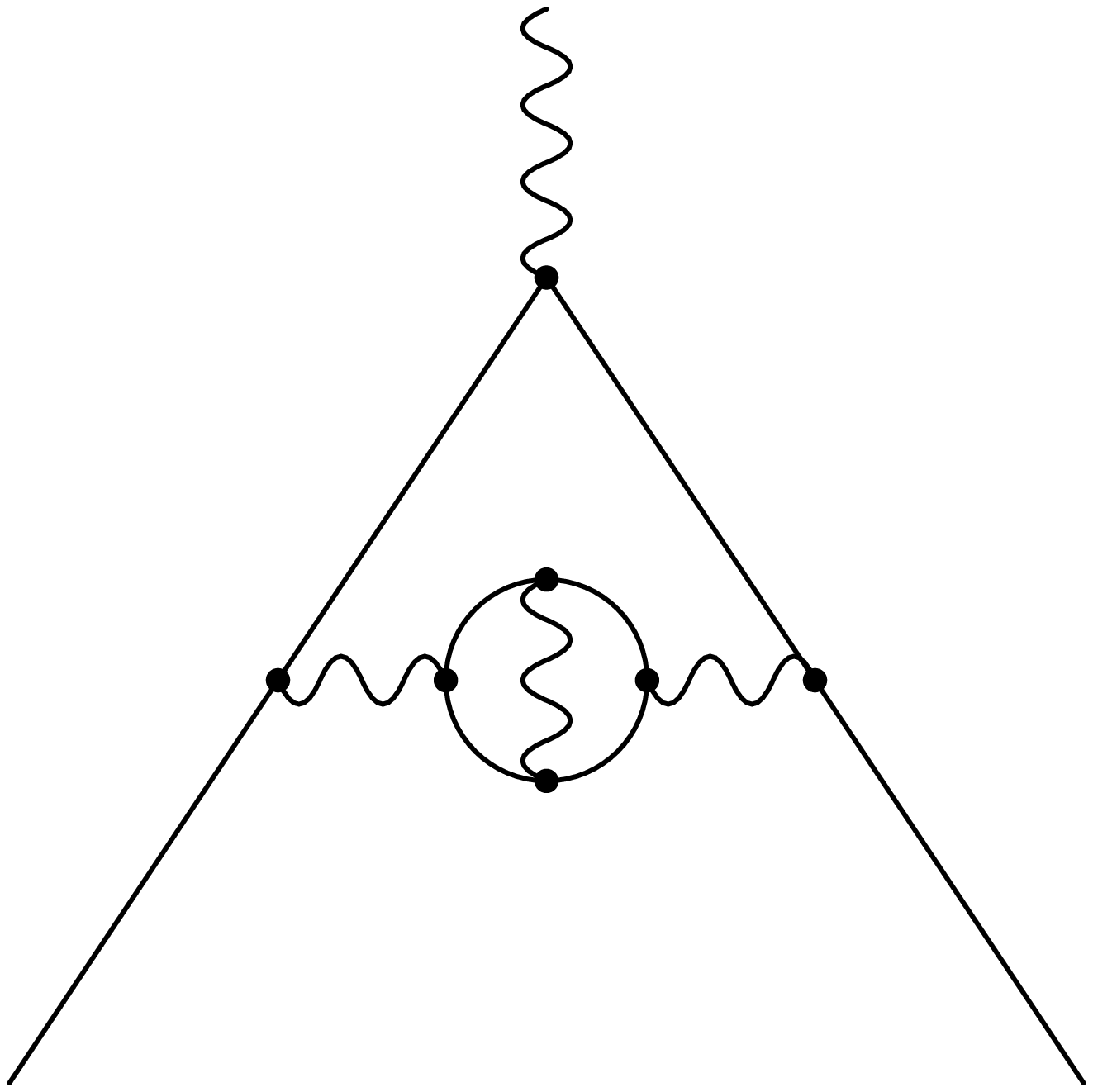, scale=0.2}}
\caption{\label{fig8}The light-by-light contribution (left) and the two-photon
  Green function (right) to the MAMM at NLO}
\end{figure}

Still the class of interpolation models based on field theories contains more
than just a useful set of functions with suitable properties. It can be used,
with some caution, for extrapolations to higher order correlation functions as
well. Indeed, the interpolation models Eqs.~(\ref{modstep},\ref{lasttech})
suffice for the calculation of the polarization-type hadronic contributions to
the MAMM related to the two-point correlator of hadronic EM currents. For the
whole computation at NLO one needs a four-point correlator of hadronic
currents. It appears in two instances: as a contribution to the light-by-light
graph and a two-photon Green function (see Fig.~\ref{fig8}). For the
contributions to the two-photon Green function the relevant projection of the
four-point correlator depends on one external momentum and the above
discussion is applicable theoretically as the interpolation function of one
complex variable is necessary. There is no experimental data to fix the IR
scale though. For the light-by-light graph the projection of the four-point
correlator relevant for the MAMM calculation is a function of two independent
four-momenta $k_1,k_2$.
The scalar form factors $g_4(k_1^2,k_2^2,(k_1-k_2)^2)$ of the projection of
the tensor four-point correlator relevant for the MAMM computation are given
by
\begin{eqnarray}
\label{g4f}
i^2\int \langle T x^\omega j_\mu^{had}(x)j_\nu^{had}(y)
j_\tau^{had}(z)j_\sigma^{had}(0) \rangle  e^{ik_1 y+ik_2 z}dx\,dy\,dz
\nonumber \\
=\sum_i{\cal T}^{\omega\,i}_{\mu\nu\tau\sigma}(k_1,k_2)~g_4^i
\left(k_1^2,k_2^2,(k_1-k_2)^2\right).
\end{eqnarray}
The functions $g_4^i$ are functions of the three scalar variables $k_1^2$,
$k_2^2$, and $(k_1-k_2)^2$. Here
${\cal T}^{\omega\,i}_{\mu\nu\tau\sigma}(k_1,k_2)$ are tensor structures
which are polynomials in the four-momenta $k_1,k_2$ and the metric tensor
$g_{\mu\nu}$. The form factors $g_4^i$ depend on the IR hadronic scale $M_h$
which is related to the experimental masses of light hadrons (pions). It may
effectively differ from the scale emerging in the two-point correlator. These
IR scales cannot be found theoretically in QCD as the regime of strong
coupling is not treatable in perturbation theory. They can perhaps be
estimated in some nonperturbative approach as, for example, the lattice
approximation. The scalar form factors $g_4^i(k_1^2,k_2^2,(k_1-k_2)^2;M_h)$
are integrated over $k_1,k_2$ with the weight functions $w(k_1,k_2,m)$ to give
a contribution to the MAMM in a full analogy with the two-point correlator
\begin{equation}
\label{g4amp}
\sum_i\int g_4^i\left(k_1^2,k_2^2,(k_1-k_2)^2;M_h\right)w^i(k_1,k_2,m)
  d^4 k_1 d^4 k_2.
\end{equation}
The weight functions $w^i(k_1,k_2,m)$ are generated by perturbative diagrams
and depend on integration momenta $k_1$, $k_2$ in the loops and on the muon
mass $m$ (more precisely, they depend also on the muon momentum $p$ which is
taken on the muon mass-shell $p^2=m^2$). To the leading order of expansion in
$m/M_h$ the integration in explicit models reduces to vacuum three loop
bubbles which can be done analytically~\cite{Broadhurst:1999rz}. 

The calculation of the integral in Eq.~(\ref{g4amp}) within an hadronization
procedure requires one to establish the analytic properties of a given form
factor $g_4^i(z_1,z_2,z_3;M_h)$ as a function of three complex variables $z_i$.
Also one has to saturate the singularities with the contributions of hadrons
in full analogy with the two-point correlator. Then the integral in
Eq.~(\ref{g4amp}) is calculated through the discontinuities on the
singularities. In other words, one can use the dispersion representation for
the form factors $g_4^i(k_1^2,k_2^2,(k_1-k_2)^2;M_h)$ with some spectral
density and then integrate over the momenta $k_1$, $k_2$ explicitly. At this
point one is left with an integral over the whole physical spectrum for the
form factors with kernels obtained after such integration. Note that this is
close to the way how the actual integration of the light-by-light diagram was
done analytically for leptons where the physical spectrum can be computed in
perturbation theory~\cite{lblanal}.

While the physical spectrum at low energies cannot be computed in QCD
perturbation theory point-wise, the integrals over the hadronic spectrum for
the three-point functions are quite accurately calculated in the sum rule
approach based on duality between hadron and quark-gluon
contributions~\cite{Nesterenko:1982gc}. As one needs only integrals of the
functions $g_4^i(k_1^2,k_2^2,(k_1-k_2)^2;M_h)$ one can avoid constructing
these functions point-wise. Assuming general requirements of smoothness one
can use an interpolation function in the Euclidean domain. Recently it has
been argued that the neutral pion contribution to the amplitude
Eq.~(\ref{g4amp}) does not give a singular dominant contribution
by itself in the kinematical region relevant for the MAMM calculation and can
be treated within the duality approach~\cite{anom1} (see also
Ref.~\cite{ope}). This implies that the integration over the Euclidean domain
properly takes into account singularities of the $g_4^i$-functions related
both to the pion contribution and two-particle cuts given by the two-pion
states. Thereby all contributions at low energies of the integrand (spectrum)
are taken into account, for example, the $\rho$ meson. The integral of the
spectrum can therefore be approximated by an integral of some interpolation
function with an appropriate IR parameter. As the IR scale is set by the pion
mass $m_\pi$ one would expect a numerical value of the IR parameter to be of
the same order of magnitude. In contrast to the two-point correlator
$\Pi^{\rm had}(q^2)$ the $g_4^i$-functions related to the four-point
correlator in Eq.~(\ref{g4f}) are complicated functions of three complex
variables which makes it difficult to find an appropriate candidate for an
interpolation function obeying all general requirements of field theory. A
simple way to find such an appropriate interpolation is to use a field theory
for constructing the spectrum. The use of a free fermionic theory is just the
way one can generate the interpolation function having the necessary
properties. Indeed, it does not violate any general principles (gauge
invariance, discrete symmetries, analyticity, unitarity, \dots) as the finite
order perturbative field theory is about the only realistic field theory that
is known to obey the general principles. One can consider such an
interpolation as an efficient adaptive integration procedure like the spline
technique or Monte Carlo routine VEGAS~\cite{Lepage:1980dq}. It is not a
full-scale approximation of QCD at low energies that would allow one to
compute the exclusive characteristics of hadrons, i.e.\ to compute the
functions $g_4^i(k_1^2,k_2^2,(k_1-k_2)^2;M_h)$ point-wise. It is a way to find
integrals of the special type in Eq.~(\ref{g4amp}) using duality between
quarks and hadrons. Note that the duality idea between $s$-channel resonances
and Regge trajectories was very useful in description of hadron scattering
before the invention of QCD. Thus, many model interpolations are possible in
the leading order and give rather accurate results. However, in the higher
orders of EM perturbation theory there are strict constraints on the
interpolation functions that one can use. Indeed, it is difficult to
generalize the models given in Eqs.~(\ref{expint},\ref{bwinter}) to higher
orders of perturbation theory. The first model is rather artificial indeed and
is not literally realized in any field theory while the second requires to
work out the full-scale hadronization of QCD for the four-point correlator
already at NLO. This is technically difficult as the data-based approach
shows~\cite{Bijnens:1996xf,Prades:2001zv}. At the same time the model
given in Eq.~(\ref{lasttech}) for the two-point correlator can immediately be
extended to any order of perturbation theory in the electromagnetic
interaction as it is a free fermion theory with $m_q$ being an IR regulator.
One also knows that this model is accurate for the LO hadronic contributions
with the IR scale taken from the data. The IR parameter effectively estimates
the distance from the Euclidean domain of integration to the physical
singularities of the hadronic correlation functions. The key physical point of
the quantitative analysis of the hadronic contributions to the MAMM within
this approach is that the same parameter $m_{\rm eff}$, or $m_q$, enters both
two-point and four-point correlators. This is, of course, an assumption which
is based on the observation that the numerical value of this parameter is
close to the pion mass. The absence of the neutral pion pole in the $g_4^i$
functions is essential for such an assumption to be valid. The result of the
analysis of the total next-to-leading order hadronic contributions to
the MAMM based on the fermionic interpolation with the same value of the
effective IR scale $M_h=m_q$ for the two-point (polarization type) and
four-point (light-by-light) correlators~\cite{anom1} 
\begin{equation}
\label{finalres}
a_\mu^{\rm mod}({\rm NLO})=(85\pm 20)\times 10^{-11}
\end{equation}
agrees with the experimental value from Eq.~(\ref{futureNLOexp}). Note that
this result includes the explicit contribution of the EM correction to the
polarization function at the NLO, i.e.\ the basic normalization quantity
obtained from the data as given in Eq.~(\ref{normLO}) is supposed not to
contain this type of contributions. This can be a rather clumsy arrangement
from the experimental point of view but it is more definite concerning the
theoretical quantities. Numerically the difference is well within error bars
for the LO contribution though.

If the scalar field theory of charged pions given by the Lagrangian
\begin{equation}
L_{\pi}=|D_\mu\pi|^2-m_\pi^2 \pi^2,\quad D_\mu=\partial_\mu-i e A_\mu
\end{equation}
is used to generate the four-point correlator of hadronic currents and thereby
the form factors $g_4^i$ that enter as integrands in Eq.~(\ref{g4amp}) then
the IR scale is explicitly identified with the pion mass and can be taken from
experiment. In this case one has to add the contribution of higher resonances
to satisfy the duality constraints at large energies which would make this
approach equivalent to the one based on explicit hadronization. The free
fermionic theory with the QCD arrangement of quantum numbers is exceptional in
this sense as it automatically complies with the duality constraints at large
energies. One, of course, should remember that this is a model and its large
energy behavior is accurate only up to higher order QCD corrections. For the
applications of interest this is inessential. A technical advantage of the
fermionic theory is that the analytical results for the MAMM at NLO are
available. For the scalar theory numerical results are available at present
although the calculations can in principle be done analytically as well since
all necessary master integrals have been found~\cite{lblanal,Laporta}. They
have already been used in three loop calculations. According to the
hadronization picture the contributions of the fermionic interpolation
functions at low energies should be substituted by the pionic ones. In the
pure fermionic model with a small effective mass the replacement of the
hadronic contributions by the model ones is effectively done at rather low
energies which makes the separate contribution of pions small or even
vanishing.

\section{Summary and conclusions}
A parameterization of the photon vacuum polarization function related to the
light hadronic modes is described in the Euclidean domain. The model contains
a single parameter which is fixed from the experimental result for the LO
hadronic contribution to the MAMM. The model describes the NLO hadronic
contributions of the vacuum polarization-type in agreement with existing
estimates. The calculation of the total NLO hadronic contribution to the MAMM
in a closely related model based on the fermionic interpolation of correlators
of hadronic electromagnetic currents is also discussed.

\vspace{5mm}
\noindent
{\large\bf Acknowledgments}\\[2mm]
We thank K.~Chetyrkin for his interest in the work and useful discussions.
This work is partially supported by the Russian Fund for Basic Research 
under contract 99-01-00091 and 01-02-16171 and by the INTAS grant. 
A.A.P. as well as S.G. gratefully acknowledge grants given by the Deutsche
Forschungsgemeinschaft.

\vspace{5mm}
\noindent
{\large \bf Note added}\\[1mm]
Recently the authors of Ref.~\cite{pseudo} have updated their result
concerning the neutral pion contribution to the light-by-light diagram. While
the magnitude of the numerical value remains the same, the sign of the
contribution has been changed~\cite{Hayakawa:2001bb}. This change brings the
explicit hadron-based result for the light-by-light contribution rather
close to that obtained within the duality approach of Ref.~\cite{anom1} which 
has further been investigated and developed in the present paper (originally
posted at the hep-ph ArXiv as hep-ph/0111206). Thus, the present theoretical
result for the muon anomalous magnetic moment agrees with the current
experimental value.

\end{document}